\documentclass[twocolumn]{aastex61}
\usepackage{rotating}
\shorttitle{}
\shortauthors{Shen et al.}
\newcommand{\speed}[1]{#1 km~s${}^{-1}$}
\newcommand{\accel}[1]{#1 m~s${}^{-2}$}
\newcommand{\nfig}[1]{Figure~\ref{#1}}
\newcommand{\rsun}[1]{${#1}\,R_\odot$}
\newcommand{\tbl}[1]{Table~\ref{#1}}
\begin{document}

\title{A quasi-periodic fast-propagating magnetosonic wave associated with the eruption of a magnetic flux rope}
\correspondingauthor{Yuandeng Shen}
\email{ydshen@ynao.ac.cn}
\author{Yuandeng Shen}
\affiliation{Yunnan Observatories, Chinese Academy of Sciences,  Kunming, 650216, China}
\affiliation{Center for Astronomical Mega-Science, Chinese Academy of Sciences, Beijing, 100012, China}
\author{Yu Liu}
\affiliation{Yunnan Observatories, Chinese Academy of Sciences,  Kunming, 650216, China}
\author{Tengfei Song}
\affiliation{Yunnan Observatories, Chinese Academy of Sciences,  Kunming, 650216, China}
\author{Zhanjun Tian}
\affiliation{Yunnan Observatories, Chinese Academy of Sciences,  Kunming, 650216, China}

\begin{abstract}
Using high temporal and high spatial resolution observations taken by the Atmospheric Imaging Assembly onboard the {\em Solar Dynamics Observatory}, we present the detailed observational analysis of a high quality quasi-periodic fast-propagating (QFP) magnetosonic wave that was associated with the eruption of a magnetic flux rope and a {\em GOES} C5.0 flare. For the first time, we find that the QFP wave lasted during the entire flare lifetime rather than only the rising phase of the accompanying flare as reported in previous studies. In addition, the propagation of the different parts of the wave train showed different kinematics and morphologies. For the southern (northern) part, the speed, duration, intensity variation are about \speed{875 $\pm$ 29 (1485 $\pm$ 233)}, 45 (60) minutes, and 4\% (2\%), and the pronounced periods of them are $106 \pm 12$ and $160 \pm 18$ ($75 \pm 10$ and $120 \pm 16$) seconds, respectively. It is interesting that the northern part of the wave train showed obvious refraction effect when they pass through a region of strong magnetic field. Periodicity analysis result indicates that all the periods of the QFP wave can be found in the period spectrum of the accompanying flare, suggesting their common physical origin. We propose that the quasi-periodic nonlinear magnetohydrodynamics process in the magnetic reconnection that produces the accompanying flare should be important for exciting of QFP wave, and the different magnetic distribution along different paths can account for the different speeds and morphology evolution of the wave fronts.

\end{abstract}

\section{Introduction}
Magnetohydrodynamics (MHD) waves have been observed for many years in the magnetic solar atmosphere, and the measured parameters of waves are used to diagnose the basic coronal property that is very important for understanding various solar phenomena \citep{nakariakov05b,shen14a,shen14b,long17a}. In recent decades, solar physicists have found different kinds of MHD waves and have achieved many important advances by using high temporal and spatial resolution observations. However, the reports on quasi-periodic fast-mode magneticsonic waves are still very scarce due to the low resolution data in the past, besides the observations of global shocks such as coronal extreme-ultraviolet (EUV) waves and chromospheric Moreton waves \citep[e.g.,][]{liu14,warmuth15,long17b,zong17,krause18}. For example, \cite{williams02} reported a possible detection of quasi-periodic fast wave along coronal loops, that has a period of 6 second and a phase speed of \speed{2100}. Besides, \cite{verwichte05} reported a propagating fast magnetosonic kink wave in a post-flare supra-arcade, whose speed and period are in the ranges of \speed{200 -- 700} and 90 -- 220 second, respectively. 

As an important new finding of the Atmospheric Imaging Assembly \citep[AIA;][]{lemen12} onboard the {\em Solar Dynamics Observatory} \citep[{\em SDO};][]{pesnel12}, quasi-periodic fast-propagating (QFP) magnetosonic waves have attracted a lot of attention since the discovery \citep{liu10}. The first detailed case analysis was performed by \cite{liu11}. They found that arc-shaped QFP wave fronts successively emanate near the kernel of the accompanying flare and propagate at a speed of \speed{2200} along both open and closed coronal loops, and the wave shares some common periods with the accompanying pulsation flare. Therefore, the authors conclude that the observed QFP wave and the periodicity of the companying pulsation flare should be triggered by a common origin. Whereafter, \cite{shen12a} and \cite{shen13b} analyzed another two QFP wave events. They not only confirmed that common periods are simultaneously existed in QFP waves and the accompanying flares, but also identified some additional periods in the waves that can not be found in the period spectrum of the accompanying flares. Since these unmatched periods are similar to the photospheric oscillations, the authors proposed that these long periods in the QFP waves are possibly caused by the leakage of photosphere pressure-driven oscillations. \cite{yuan13} found that the QFP wave reported in \cite{shen12a} was composed of three distinct sub-QFP waves, which have different periods, wavelengths, and amplitudes. It is interesting that each sub-QFP wave was associated with a small radio burst, which suggests that the excitation of a QFP wave should be tightly related to the non-linear energy releasing process in the magnetic reconnection process that produces the flare. \cite{goddard16} observed quasi-periodic radio bursts during a QPF wave event that was associated with a large-scale coronal mass ejection (CME), and they suggested that these radio bursts are possibly caused by the interaction between propagating QFP wave fronts and the leading edge of the associated CME. Recently, \cite{kumar17} reported that the observed QFP wave fronts are generated during the quasi-periodic magnetic reconnection near the null point as described in the magnetic breakout model \citep[e.g.,][]{antiochos99,shen12b}. Although QFP waves are different from global coronal EUV waves that are thought to be driven by CMEs \citep[e.g.,][]{shen12c,shen12d,shen13a,shen17}, many studies have indicated that the two types of coronal waves are closely related to each other \citep{liu10,liu12,shen12a,shen13b}. Similar case studies on QFP waves have been documented in literature \citep{zhang15,kumar15a,kumar15,qu17}, and the common characteristics of QFP waves based on previously published events have been summarized in \cite{liu14}.

In addition to observations, numerical experiments are also performed timely to investigate the physics nature of QFP waves. \cite{ofman11} firstly presented the three-dimensional MHD modeling of the QPF wave event that was reported by \cite{liu11}. They successfully generated QFP wave fronts whose physical properties are similar to observational results. Based on the simulation, the authors identified that the observed QFP wave fronts are fast magnetosonic waves in a funnel-shaped magnetic waveguide. With two-dimensional MHD modeling approach, \cite{pascoe13} and \cite{pascoe14} also generated QFP wave fronts in and outside a field-aligned density funnel magnetic structure, and they proposed that the wave fronts outside the magnetic structure might account for the observed wave fronts. \cite{nistico14} further compared their observational results with the model presented by \cite{pascoe13} and concluded that QFP waves can be generated by a localized impulsive energy release through dispersive evolution. \cite{qu17} also reported a QFP wave that showed leakage of wave fronts from the guiding coronal loops, and the authors proposed that their observation is possibly in agreement with the finding of \citep{pascoe13}. Some authors considered that QFP waves are possibly caused by periodic processes in magnetic reconnections. For example, \cite{yang15} showed that QFP wave fronts can be excited by the collision of outward moving plasmoids in reconnection current sheet with the ambient magnetic fields. \cite{takasao16} found that QFP waves can be spontaneously excited by the backflow of reconnection outflow above the region of flaring loops. In addition, oscillatory magnetic reconnection can also naturally produces quasi-periodic pulsations (QPPs) and QFP wave \citep[e.g.,][]{mclaughlin12a,mclaughlin12b,thurgood17}. Other theoretical and simulation works on the driving and dispersion properties of QFP waves can be found in articles \citep[e.g.,][]{roberts83,roberts84,shestov15,oliver15,pascoe16,yu16,yuan16}. In addition, since QFP waves often share similar periods with the accompanying pulsation flares that have been studied for many decades, one can draw on the experience of QPPs in flares and find some clues to understand the excitation mechanism of QFP waves \citep[e.g.,][]{nakariakov09,van16,li15a,li17a,li17b,li17c}.

So far, only a few QFP wave events are reported in literature, and the excitation mechanism and evolution processes are still unclear. Therefore, more case studies based on high temporal and spatial resolution data are needed to understand the physics in QFP waves. The present paper presents an imaging observations of a QFP wave in which more that 20 wave fronts can be clearly identified and the wave train lasted for about one hour. It is found that the wave train showed different physical characteristics along different paths, which manifest the different physical properties of the guiding fields. In addition, the lifetime of the present QFP wave is much longer that those have been reported in previous studies. It is found that the present QFP wave lasted during the entire flare process, which is different from previous cases which only exist during the rising phase of the accompanying flare. Instruments and observations are briefly introduced in Section 2, analysis results are presented in Section 3, conclusions and discussions are given in the last section.

\section{Observations and Instruments}
The present QFP wave event was observed by AIA onboard the {\em SDO} on 2014 March 23. The full-disk images are taken by the AIA instrument, which images the Sun up to \rsun{1.3} in seven EUV and three UV-visible wavelength bands. The pixel size resolution of the AIA EUV and UV images are of 0\arcsec.6, and their temporal resolutions are 12 and 24 second, respectively. The line-of-sight (LOS) magnetograms used in this paper were taken by the Helioseismic and Magnetic Imager \citep[HMI;][]{schou12} onboard the {\em SDO}. The temporal resolution of HMI LOS magentograms is 45 second, and the measurement precision is 10 Gauss. We also use the soft X-ray 1 -- 8 \AA\ flux is recorded by the {\em Geostationary Operational Environmental Satellite} ({\em GOES}), and the hard X-ray fluxes taken by the {\em Reuven Ramaty High Energy Solar Spectroscopic Imager} \citep[RHESSI;][]{lin02}, and the Nobeyama radio flux at 1 GHz to analyze the periodicity of the accompanying pulsation flare. The temporal resolutions of the Nobeyama radio, the {\em GOES} soft X-ray,  and {\em RHESSI} hard X-ray fluxes are of 44 millisecond, 1 minute and 4 second, respectively. All images used in this paper are differentially rotated to the reference time of 03:30:00 UT on March 23, and the solar north is up, west to the right.

\section{Observational Results}
On 2014 March 23, a {\em GOES} C5.0 flare occurred in NOAA active region AR12014, whose start, peak and end times are 03:05:00, 03:48:00, and 04:30:00 UT, respectively. Right before the start of this flare, another small flare of {\em GOES} C3.1 class was also detected. The start, peak, and end times of this small flare are 02:27:00, 02:37:00, and 02:45:00 UT, respectively. During the entire lifetime of the main C5.0 flare, it is observed that multiple arc-shaped wave fronts successfully emanated from the flare kernel and propagated along funnel-shaped open coronal loops rooted in the active region. In addition, it is observed that the QFP wave was associated with the eruption of a magnetic flux rope from the active region and a halo CME observed in the LASCO coronagraphs. According to the measurement of Coordinated Data Analysis Workshops (CDAW) CME catalog\footnote{\url{https://cdaw.gsfc.nasa.gov/CME_list}}, the first appearance of the CME in the field of view (FOV) of LASCO C2 was at 03:36:00 UT, and the linear speed and the acceleration of the CME were \speed{820} and \accel{2.3}, respectively. Considering the temporal relationship between the flare and the CME, it can be derived that the start time of the CME from the eruption source region should be at about 03:10:00 UT, which suggests that the CME should be related to the main C5.0 flare recorded by {\em GOES}.

\begin{figure*}[thbp]
\epsscale{1}
\figurenum{1}
\plotone{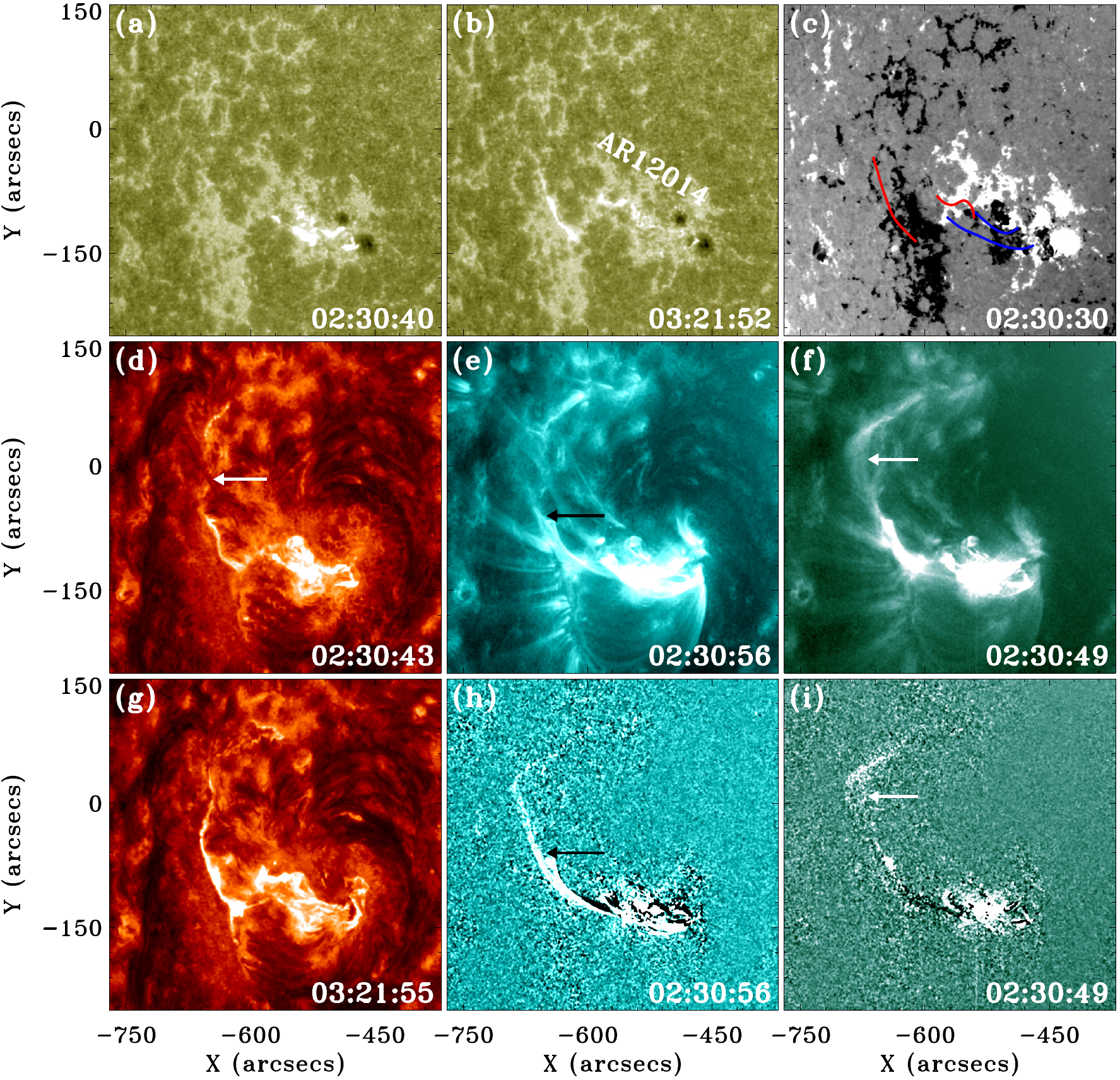}
\caption{An overview of the eruption source region. Panels (a) and (b) are AIA 1600 \AA\ images. Panel (c) is an HMI LOS magnetogram, in which the red and blue lines mark the flare ribbons determined from panels (a) and (b), respectively. Panels (d) and (g) are AIA 304 \AA\ images. Panels (e) and (f) are AIA 131 and 94 \AA\ images, and panels (h) and (i) are the AIA 131 and 94 \AA\ running difference images, respectively. The arrow in panel (d) indicates the rising filament, while the arrows in panels (e), (f), (h), and (i) point to the erupting flux rope.
\label{fig1}}
\end{figure*}

\nfig{fig1} shows the source region of the flares and the erupting flux rope with AIA multi-wavelength images. The entire eruption process was composed of two flares. The first C3.1 flare occurred close to the two sunspots, and the pair of flare ribbons can well be identified in the AIA 1600 \AA\ images (see \nfig{fig1} (a) and the blue lines in \nfig{fig1} (c)). About twenty minutes after the first flare, the main C5.0 occurred on the eastern side of the sunspots. This flare was associated with the eruption of a filament in AIA 304 \AA\ images (see the white arrow in \nfig{fig1} (d)). The flare ribbons are obvious in AIA 1600 \AA\ (\nfig{fig1} (b)) and AIA 304 \AA\ (\nfig{fig1} (g)) images. The location of the two ribbons are also overlaid on the HMI LOS magnetogram as red curves in \nfig{fig1} (c). During the eruption process of the filament, a loop-like hot magnetic structure is observed in the hot AIA 131 and 94 \AA\ observations (see the arrows in \nfig{fig1} (e) and (f)), which should be a hot flux rope structure as reported in previous studies \citep[e.g.,][]{zhang12,cheng13,yan15,yan17}. The flux rope also erupted along with the eruption of the filament, and their eruption directly caused the observed halo CME in the LASCO coronagraphs.

\begin{figure*}[thbp]
\epsscale{1}
\figurenum{2}
\plotone{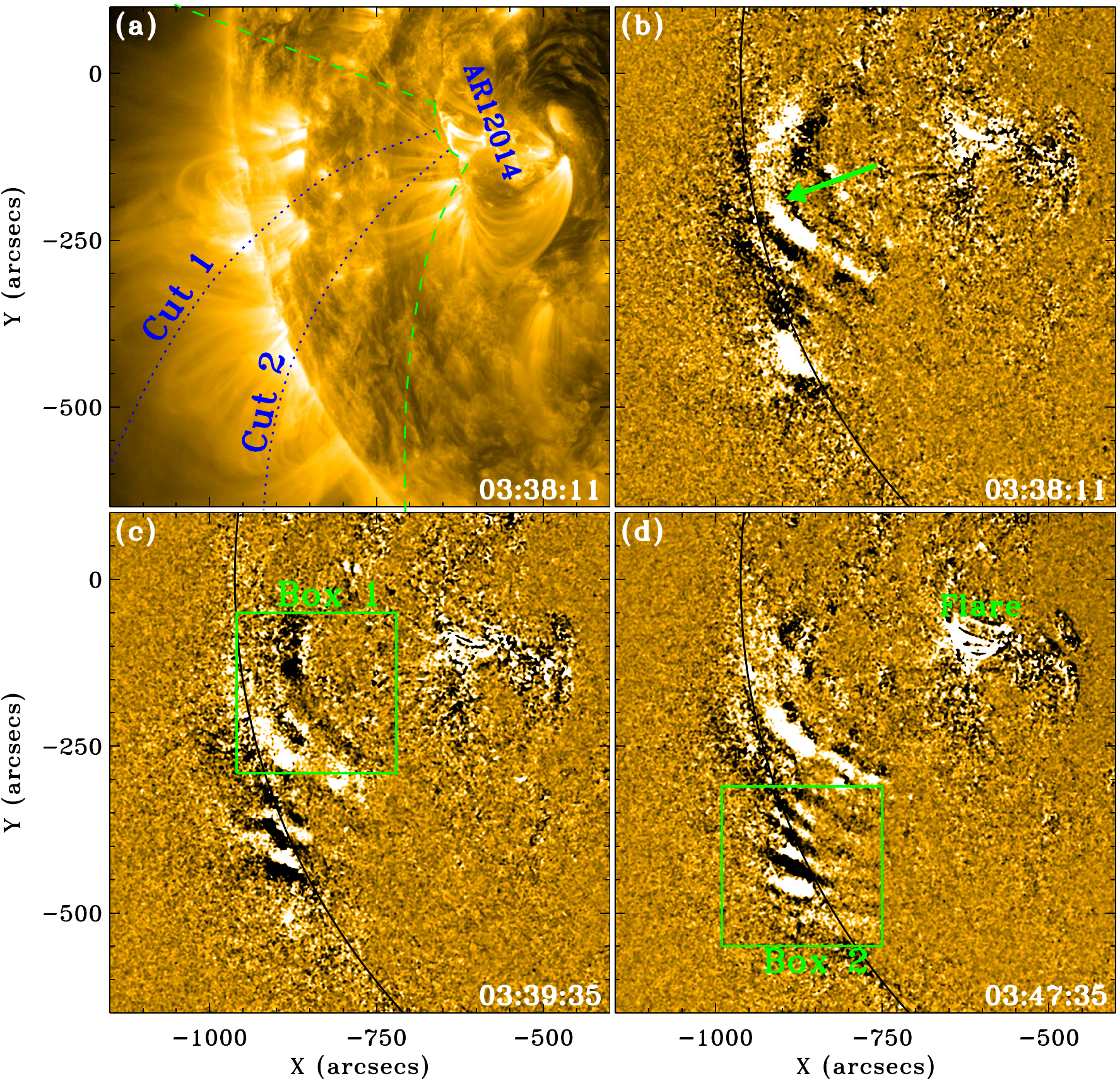}
\caption{Panel (a) is an AIA 171 \AA\ direct image. The dotted curves (Cut 1 and Cut 2) show the paths used to obtain time-distance diagrams, and the dashed green curves indicate the region where the wave train propagates. Panels (b) -- (d) are AIA 171 \AA\ running difference images show the evolution of the wave train. The arrow in panel (b) points to the position where refraction effect of the wave front is observed, and the boxes in panels (c) and (d) indicate the regions for Fourier analysis. An animation is available in the online journal.
\label{fig2}}
\end{figure*}

During the eruptions of the flares and the flux rope, multiple arc-shaped QFP wave fronts are observed in the southeast of AR12014. The bright arc-shaped wavefronts can be clearly identified in the AIA 171 \AA\ running difference images. It is noted that some faint wave signal can also be observed in the AIA 193 \AA\ running difference images. In the present paper, we mainly use the AIA 171 \AA\ observations to study the physical property of the QFP wave. An overview of the region where the wave train propagates and the evolution process of the wave fronts are displayed in \nfig{fig2}. The propagation region of the wave train is indicated by the two green dashed  curves in \nfig{fig2} (a). This region is full of funnel-shaped open coronal loops rooted at the periphery of the active region (see \nfig{fig2} (a)), which can be regarded as the waveguide of the observed QFP wave train. The evolution of the wave fronts are displayed in panels (b) -- (d) of \nfig{fig2}. It is observed that successive wave fronts continuously emanated from the periphery region of the active region and became pronounced at about 100 Mm away from the flare kernel, and the northern part of the wave fronts became boarder and more bent when they propagated to the eastern limb of the solar disk (see the green arrow in \nfig{fig2} (b)). The deformation of the wave fronts is probably due to the refraction effect when they penetrated into a magnetic region whose magnetic field strength is stronger than the ambient region. Such an phenomenon is similar to global EUV waves when they pass through a strong magnetic region such as active region \citep{shen13a}. For the southern part of the wave train, the shape of the wave fronts do not show obvious changes. The different evolution patterns of the wave train along the two different paths might manifest the different magnetic environments of the guiding coronal loops. For more detailed evolution process of the QFP wave, one can see the online animation associated with \nfig{fig2}.

\begin{figure*}[thbp]
\epsscale{1}
\figurenum{3}
\plotone{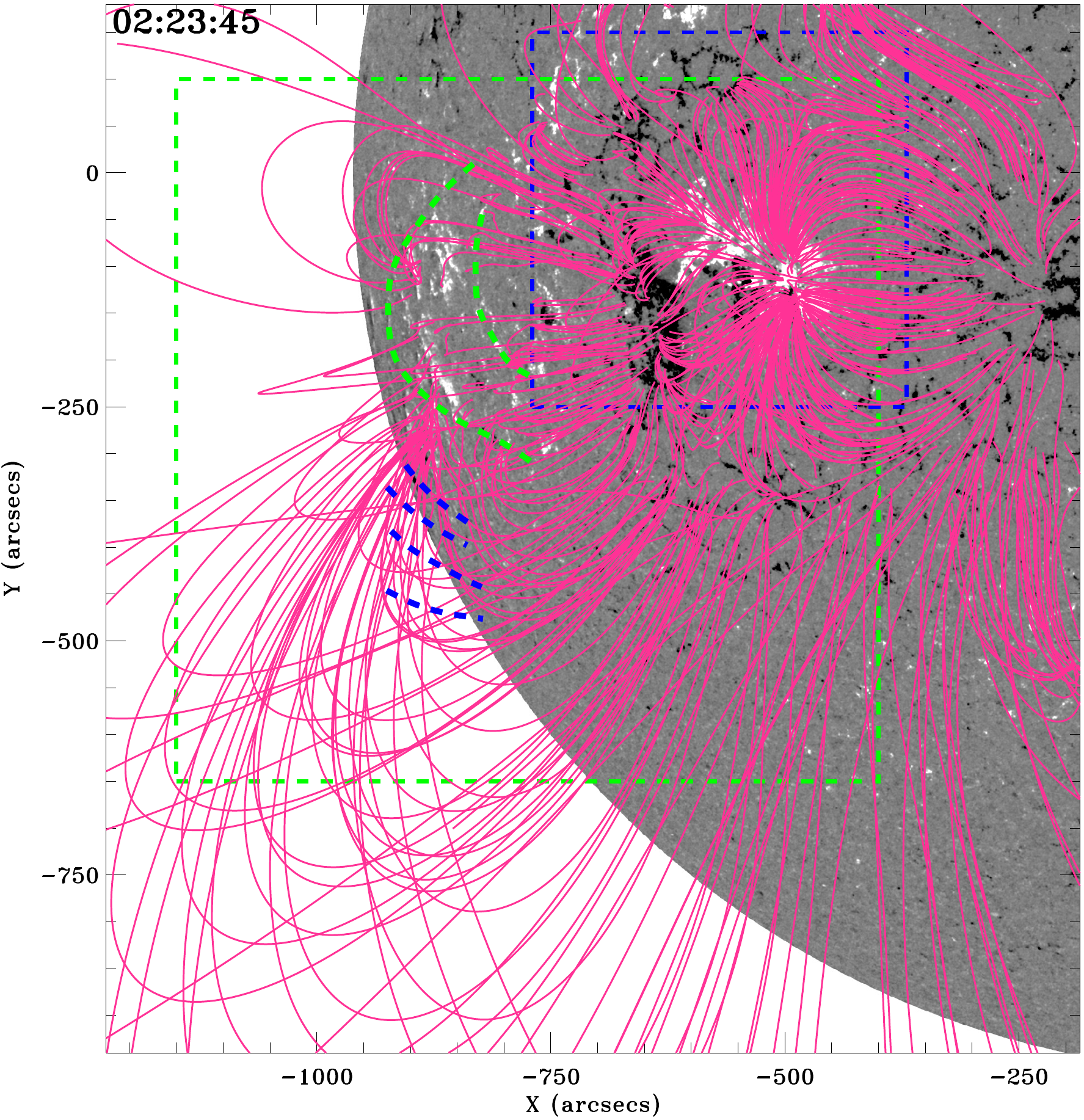}
\caption{An HMI LOS magnetogram overlaid with magnetic field lines (pink) extrapolated by using the PFSS model \citep{schrijver03}, in which the black and white patches are the regions of negative and positive magnetic polarities, respectively. The dashed blue and green boxes show the FOVs of the panels in \nfig{fig1} and \nfig{fig2}, respectively. The green and blue dashed curves mark the position of the QFP wave trains at 03:38:11 UT and 03:47:35 UT in the AIA 171 \AA\ images, respectively.
\label{fig3}}
\end{figure*}

Although the distribution of the coronal loops in the AIA 171 \AA\ observations have given us some hints about the magnetic environment of the region where the wave train propagates, we further extrapolated the three-dimensional coronal magnetic fields by using the potential field source surface (PFSS) software available in the SolarSoftware (SSW) package \citep{schrijver03}. In the PFSS model, the magnetic fields in the corona are assumed to be a potential field (current-free) and become radial at the source surface at \rsun{2.5}, and the extrapolation is based on a synoptic magnetic map that was composed of a series of consecutive HMI LOS magnetograms within a limited area around the central meridian. It should be pointed out that the extrapolated coronal field may depart from the actual solar field in magnitude and direction, due to the boundary condition and potential assumptions in the PFSS model. However, the basic topological structure and orientation should be reliable. The extrapolated magnetic field lines are overlaid on the HMI magnetogram at 02:23:45 UT as pink curves in \nfig{fig3}. In the meantime, the positions of the wave fronts for the northern part determined at 03:38:11 UT are also overlaid in \nfig{fig3} as green dashed curves, while those for the southern part determined at 03:47:35 UT are plotted as blue dashed curves. It can be seen that the magnetic fields in the eruption source region are mainly lower closed loops, while the region where the wave train propagates are in fact also closed loops, but with larger height than that in the eruption source region. The region where the northern part of the wave train propagates is a region with positive magnetic polarity that has a relatively stronger magnetic field strength than the quiet-Sun region. This can explain the refraction effect of the northern part of the wave train.  For the region where the southern part of the wave train propagates, the higher closed guiding loop can be regarded as quasi-open loop for the QFP wave, since their propagation distance is much smaller than the height of the loops. In addition, the orientation of the guiding coronal loops well indicates the propagation direction of the northern part of the wave train.

\begin{figure*}[thbp]
\epsscale{1.1}
\figurenum{4}
\plotone{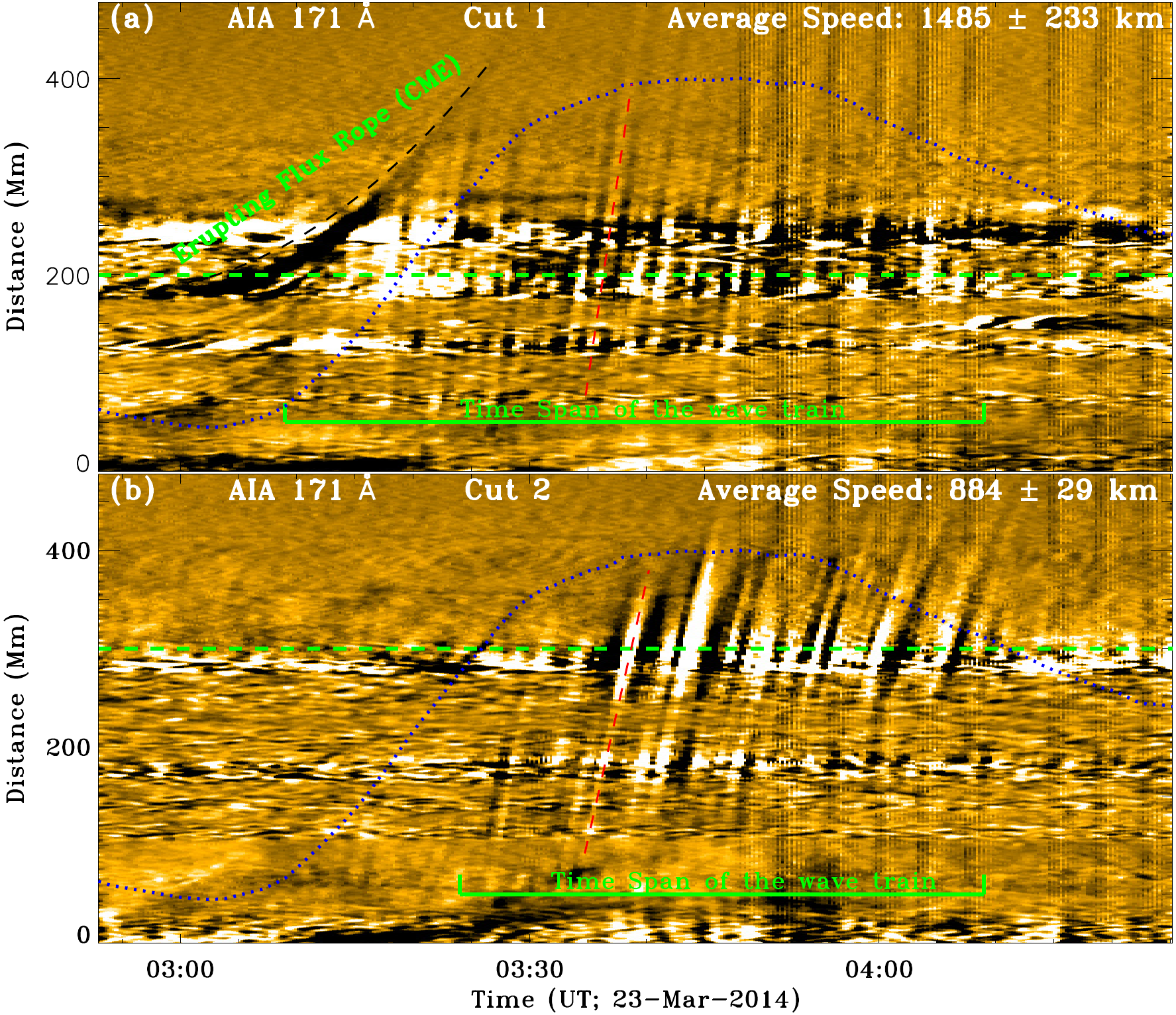}
\caption{Panels (a) and (b) are time-distance diagrams obtained from AIA 171 \AA\ running difference images along Cut 1 and Cut 2, respectively. In each panel, the horizontal green line indicates the duration of the QFP wave, the red dashed line is the linear fit to the bright ridge, and the horizontal dashed line marks the position where the intensity variation is analyzed. The slopes for the red dashed lines in the top and bottom panels are \speed{1340 and 872}, respectively. The black dashed curve in panel (a) is the second order polynomial fit to the expanding loop, and the dashed red lines are linear fit to the wave ridges. The blue dotted curves are the {\em GOES} 1 -- 8 \AA\ soft X-ray flux.
\label{fig4}}
\end{figure*}

The kinematics of the wave fronts is shown in \nfig{fig4} using time-distance diagrams along the two paths of Cut 1 and Cut 2 as shown in \nfig{fig2} (a), and these diagrams are made from AIA 171 \AA\ running difference images. Here, a running difference image is obtained by subtracting the present image by the previous one in time, and moving features can be observed clearly in running difference images. To obtain a time-distance diagram, we first obtain the one-dimensional intensity profiles along a specified path at different times, and then a two-dimensional time-distance diagram can be generated by stacking the obtained one-dimensional intensity profiles in time. Along the path of Cut 1, the erupting flux rope can be identified before the appearance of the wave train (see \nfig{fig4} (a)), which underwent a slow expanding phase at an acceleration of \accel{65}. The wave train started at about 03:08:00 UT, corresponding to the fast eruption phase of the flux rope. In the time-distance diagram, each inclined bright ridge represents the propagation of a wave front. It is clear that more than 20 wave fronts can be identified from \nfig{fig4} (a), and they appeared and disappeared at distances of about 50 Mm and  400 Mm from the origin of Cut 1, respectively. The duration of the wave train is about 60 minutes. The propagation speed of each wave front is obtained by fitting the ridge with a linear function. For example, the red dashed line in \nfig{fig4} (a) is a linear fit to a ridge, which yields the speed of the wave front, and the value is \speed{1340}. We calculate the speeds of the all wave fronts and obtain that the average speed of these wave fronts is about \speed{1485 $\pm$ 233}. Along the path of Cut 2, the appearance of the first wave front was at about 03:25:00 UT, and the duration of these wave train is about 45 minutes. With the same method, we obtain the average speed of the wave fronts along Cut 2 is about \speed{875 $\pm$ 29}, which is about 60\% of the wave speed along Cut 1. Here, it is noted that the duration of the QFP wave is almost the same with the lifetime of the flare (see the blue dotted curve in \nfig{fig4}), which is different from previous reported QFP waves \citep[e.g.,][]{liu11,shen12a,shen13b,nistico14,kumar17}. This new finding may imply more physical information about the excitation mechanism of the QFP wave.

\begin{figure*}[thbp]
\epsscale{1}
\figurenum{5}
\plotone{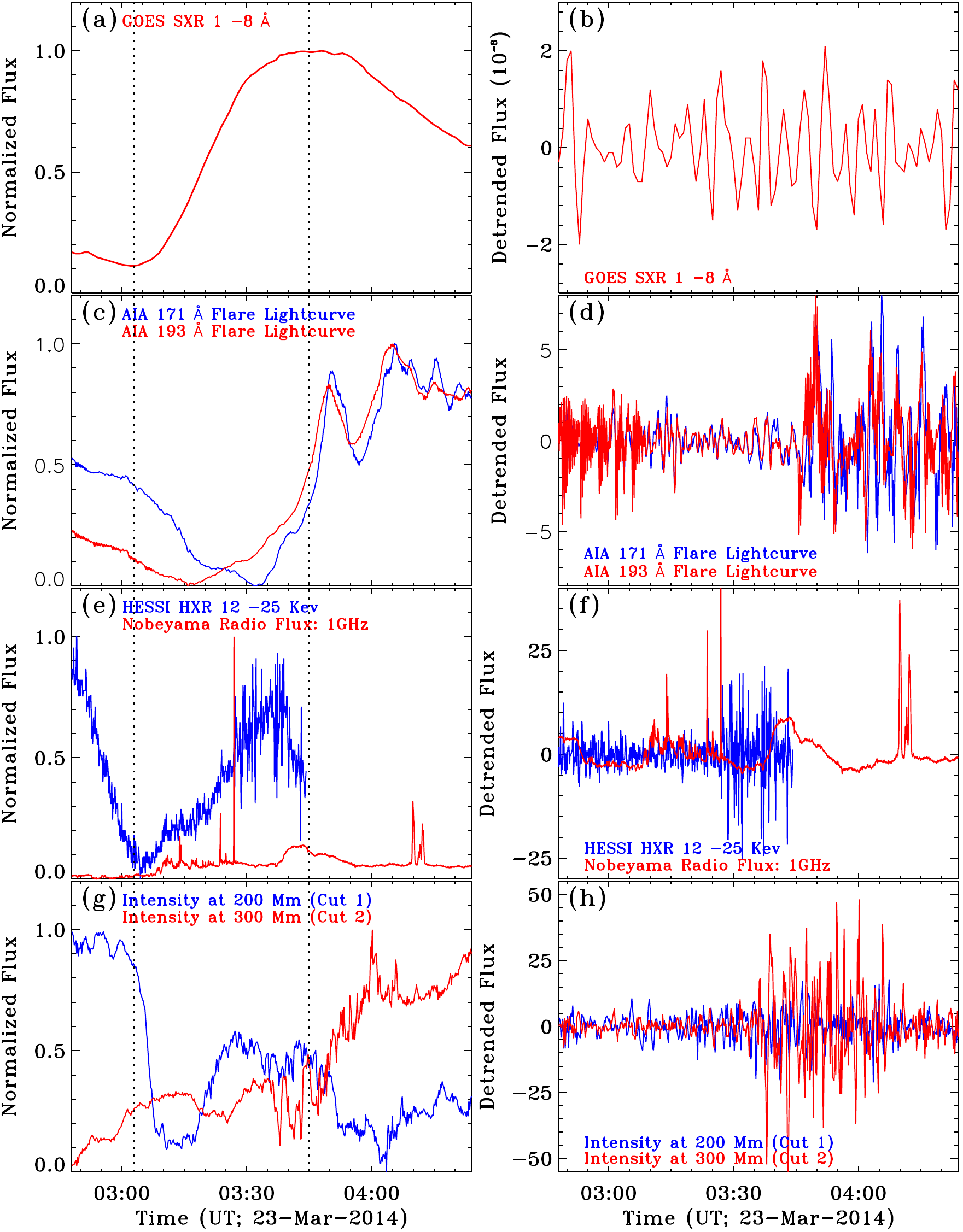}
\caption{Panels (a) is the {\em GOES} SXR 1 -- 8 \AA\ flux, while panel (b) shows the detrended curve of the derivation SXR 1 -- 8 \AA\ flux. Panel (c) shows the AIA 171 \AA\ (blue) and 193 \AA\ (red)  flare lightcurves, while panel (e) shows the {\em RHESSI} HXR count rates in the energy band (4 s integration) of 12 -- 25 keV (blue) and Nobeyama radio flux at 1 GHz (red). Panel (g) shows the intensity variations of the wave trains at the positions as shown by the horizontal dashed lines in \nfig{fig4}. The right panels show the corresponding detrended flux curves at different wavelength. The vertical dotted lines in the left column indicate the rising phase of the flare.
\label{fig5}}
\end{figure*}

Since QFP waves often share common periods with the accompanying pulsation flares, we further check the periodicity of the accompanying flare by analyzing the flare lightcurves from radio to hard X-ray (HXR) waveband. In the meantime, the intensity variations of the wave train are also extracted to analyze the periodicity of the QFP wave. In \nfig{fig5}, the left column shows the flare lightcurves (panels (a), (c), and (e)) and the intensity variations of the wave train (panel (g)). For each panel in the left column, the corresponding detrended fluxes are plotted on the right panel. It can be seen that all the variation of these fluxes in time show obvious pulsations during the lifetime of the QFP wave. Especially, the Nobeyama radio flux at 1 GHz showed several bursts during the rising phase of the main C5.0 flare, which suggests the periodic energy releasing process in the flare as reported in \cite{yuan13}. It is measured that the intensity variation amplitude are about 2\% and 4\% relative to the background intensity for the wave train along Cut 1 and Cut 2, respectively. The intensity variation of the wave train is in agreement with the first QFP wave event reported by \cite{liu11}, where the authors found that the intensity variation of the wave fronts is in the range of 1\% -- 5\%. Such a low intensity variation rate can explain why QFP waves are hard to be observed by eye in the direct AIA images (see also the left column of the online animation).

\begin{figure*}[thbp]
\epsscale{0.9}
\figurenum{6}
\plotone{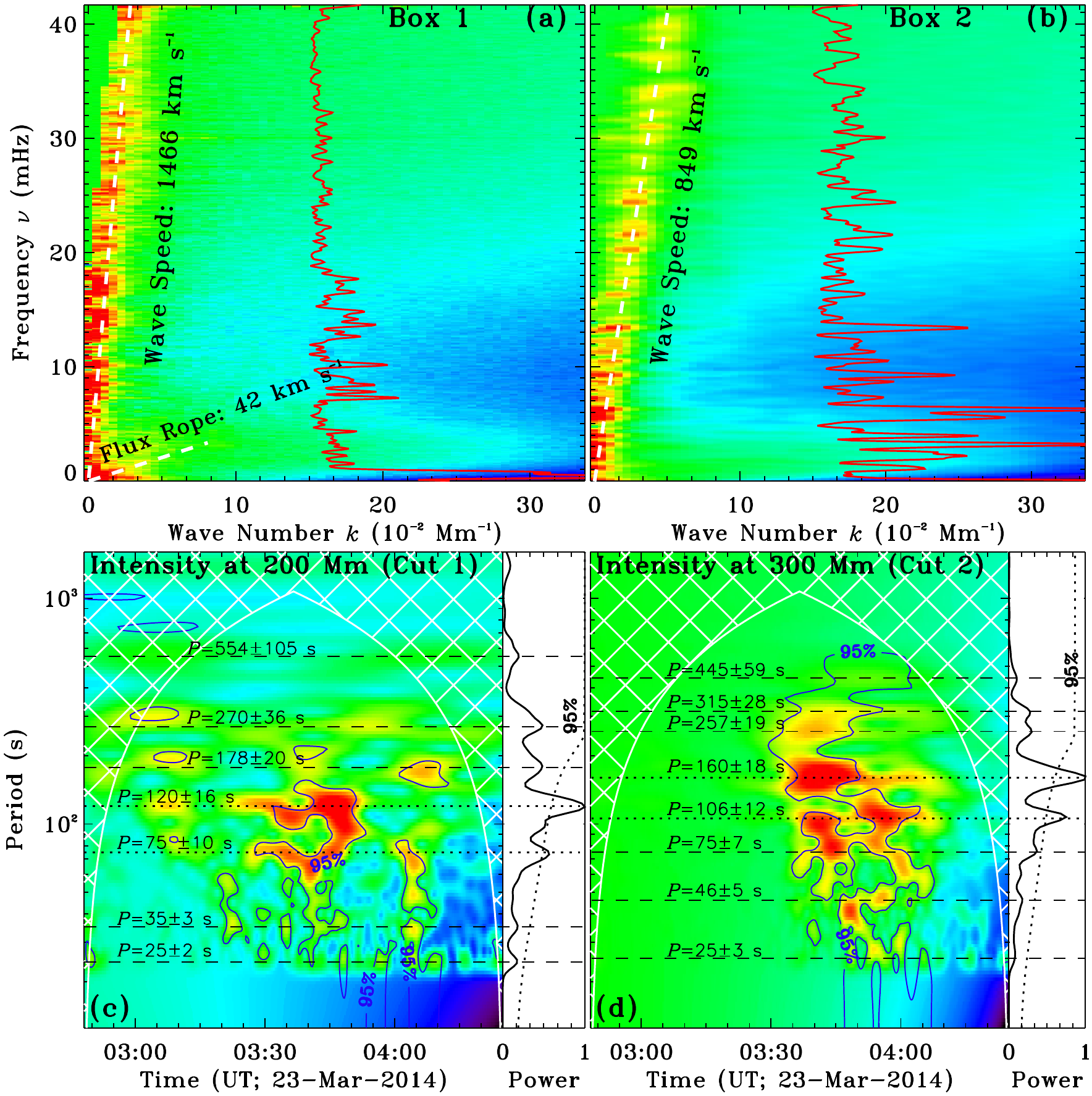}
\caption{Panels (a) and (b) show the Fourier power ($k$--$\omega$) diagrams of the three-dimensional data cube of AIA 171 \AA\ running difference images in the two box regions as shown in \nfig{fig1}, in each panel the overlaid red curve is the normalized intensity profile of the straight ridge. Panels (c) and (d) are the wavelet power maps of the detrended intensity profiles at distances of 200 Mm (Cut 1) and 300 Mm (Cut 2) from the flare, respectively. In each wavelet power map, the blue contours outline the regions where the significance is above 95\%, and the periods are highlighted by horizontal dashed lines. In addition, the corresponding normalized global power is plotted on the right, in which the dotted line indicates the 95\% significance level.
\label{fig6}}
\end{figure*}

\begin{table*}[thbp]
\centering
\caption{Periods of the QFP wave along the two different paths}
\begin{tabular}{ccccccccccc}
\hline
\hline

\multicolumn{2}{c}{\bf Distance\tablenotemark{a} (Mm)}  &\multicolumn{9}{c}{\bf Period of the wave (s)}\\
\hline
\multicolumn{2}{c}{200 (Cut 1)}   & 25$\pm$2 & 35$\pm$3 & ... & 75$\pm$10 & 120$\pm$16  & 178$\pm$20 & 270$\pm$36 & ... & 554$\pm$105 \\
\multicolumn{2}{c}{300 (Cut 2)}  &25$\pm$3 &... & 46$\pm$5 & 75$\pm$7 & 106$\pm$12  & 160$\pm$18 & 257$\pm$19 & 315$\pm$28 & 445$\pm$59 \\
\hline
\end{tabular}
\tablenotetext{a}{The distance from the flare along different paths.}
\tablecomments{The units of the distances and periods are Mm and second, respectively.}
\end{table*}
\label{tbl1}

Using the method presented in \cite{deforest04}, Fourier analysis of the QFP wave are performed in the two box regions shown in \nfig{fig2}, and the time interval is from 02:40:00 UT to 04:25:00 UT. The Fourier power maps ($k$--$\omega$ diagram) of the two regions are plotted in \nfig{fig6} (a) and (b), respectively. In each $k$--$\omega$ diagram, a steep and narrow ridge can be identified, which describes the dispersion relation and its slope represents the propagation speed of the QFP wave. With the method of linear fit to the bright ridges, it is obtained that the speeds of the two part of the wave train are of \speed{1468, and 849}, respectively. These results are in agreement with the results directly measured from the time-distance diagrams, and confirms the result that the different parts of the wave train have different propagation speeds as described above. The lower ridge in \nfig{fig6} (a) represent the expanding flux rope whose speed is about \speed{42}. On the ridge in each $k$--$\omega$ diagram, one can see many dense nodes that represent the available frequencies (periods) of the QFP wave. To better identity these nodes, the normalized intensity profile of the ridge is overlaid on the $k$--$\omega$ diagram, along which one can identify many peaks that represent the dense nodes along the bright ridges. The periodicity of the intensity variations of the wave train along the two cuts at distances of 200 Mm (Cut 1) and 300 Mm (Cut 2) are also analyzed by using the wavelet software \citep{torrence98}. This tecnique is a common method for analyzing localized variations of power within a time series, and it has been used to analyze the periodicity of various data \citep[e.g.,][]{deng13a,deng13b,deng13c}. The wavelet power maps reveal that the pronounced periods (global power significance level $> 95\%$) are about $75 \pm 10$ and $120 \pm 16$ seconds along Cut 1, while that are about $106 \pm 12$ and $160 \pm 18$ seconds along Cut 2. There are still many weak periods (global power significance level $< 95\%$) are also indicated with horizontal dashed lines in \nfig{fig6} (c) and (d). With the measured speeds and periods of the QFP wave, the corresponding wavelengths are estimated to be 19.8 and 12.4 Mm for the wave train along Cut 1, while that for the wave train along Cut 2 are 8.3 and 5.5 Mm. Obviously, the wavelength of the wave train along Cut 1 are longer than those along Cut 2. Here, because the waves in the two regions should be excited by the same physical process in the flare, the different values of the characteristic parameters of the waves in the two different regions are possibly due to the modulation of the medium in which the waves propagate. The detected periods of the QFP wave along the two paths are all listed in \tbl{tbl1}. It can be seen that some periods are simultaneously existed in the waves along the two paths (for example, periods of 25, 75, and 120 second), but some periods can only be detected along the a certain paths (for example, periods of 35, 46, and 315 second).

The periodicity of the accompanying pulsations flare is also analyzed with the wavelet software, and all periods detected from different wavelengths from radio to HXR waveband are listed in \tbl{tbl2}. For a given wavelength data, the variation of the flare lightcurve shows several periods. Meantime, some similar periods are detected simultaneously at different wavelengths. The similar periods detected from different wavelength observations are averaged and the mean values are listed in the bottom row of \tbl{tbl2}. It is found that the period of the flare ranges from 9 to 1078 second. It should be pointed out that the error of each period is determined by the full-width at half-maximum of each peak on the global power curve. By comparison the detected periods of the flare listed in \tbl{tbl2} and those of the QFP wave listed in \tbl{tbl1}, one can find that all the periods of the QFP wave can be found in the period spectrum of the accompanying flare, although some periods show a little difference rather than exactly the same. This suggests that the excitation of the QFP wave are possibly due to a common physical process with the accompanying pulsation flare, in agreement with previous studies \citep[e.g.,][]{liu11,liu12,shen12a,shen13b}.

\section{Conclusions and Discussions}
Using high temporal and high spatial observations taken by the {\em SDO}/AIA and other instruments, we present the detailed analysis results of the QFP wave occurred on 2014 March 23, which was accompanied by a {\em GOES} C5.0 flare and the eruption of a flux rope in NOAA active region AR12014. The event started from a small C3.1 around the center of the active region, which probably resulted in the destabilization of the magnetic system at first, and thus then further caused the following nearby C5.0 flare, the flux rope eruption, as well as the halo CME. During the fast eruption phase of the flux rope,  multiple arc-shaped wave fronts successfully emanated from the periphery of the active region and became pronounced at about 100 Mm away from the flare kernel. It is noted that the lifetime of this QFP wave is about an hour, and its time interval was nearly the same with entire duration of the accompanying C5.0 flare. This result is different from previous events that have been reported in literatures \citep[e.g.,][]{liu11,shen12a,shen13b,nistico14,kumar17}, where the QFP waves are only existed during the rising phase of the accompanying flare.  Based on the AIA direct imaging observations, it can be identified that the propagation of the successive wave fronts are along funnel-shaped open coronal loops rooted in the periphery of the active region, as described in previous studies \citep[e.g.,][]{liu11,shen12a}. However, the three-dimensional coronal field extrapolated based on the PFSS model suggests that the guiding coronal loops are actually closed loops rooted in the active region. Since the propagation distance of the wave train is much smaller than the height of the guiding loops, the latter can be regarded as quasi-open fields for the propagating wave fronts, and this magnetic topology of the guiding fields can naturally explain the observed funnel-shaped guiding loops based on the AIA imaging observations.

The propagation of the QFP wave showed different kinematics and morphologies for the northern and southern wave parts. For the northern part of the wave train, the speed, duration, intensity variation are about \speed{1485 $\pm$ 233}, 60 minutes, and 2\%, respectively. The pronounced periods are $75 \pm 10$ and $120 \pm 16$ seconds, and the corresponding wavelengths are 19.8 and 12.4 Mm. In the meantime,  the speed, duration, intensity variation for the southern part of the wave train are about \speed{875 $\pm$ 29}, 45 minutes, and 4\%, respectively. The pronounced periods are about $106 \pm 12$ and $160 \pm 18$ seconds, corresponding to a wavelength of 8.3 and 5.5 Mm, respectively. For each part of the wave train, we separately generated the Fourier $k$--$\omega$ diagram based on the three-dimensional data cube in the region where the wave propagates. It is found that each $k$--$\omega$ diagram shows a bright ridge with many dense nodes passing through the frequency origin, which describes the dispersion relation of the QFP wave, and the dense nodes represent the possible frequencies in the the wave. We also measured the propagation speeds of the two parts of the wave train based on the Fourier $k$--$\omega$ diagram. It is obtained that the speeds for the two parts are similar to those measured from the direct imaging observations, and their values are \speed{1468 and 849}, respectively. Therefore, our measurement results about the wave speeds should be reliable. In addition, the northern part of wave fronts showed obvious morphology changes during the propagation. It is observed that the wave fronts become more and more bend during the propagation. We propose that this phenomenon could be explained as the refraction effect of the QFP wave, similar to what has been reported in global EUV waves \citep[][]{shen13a}. Both the photospheric magnetic field and the extrapolated coronal field suggest that the northern part of the wave train pass through a region of strong magnetic field, but the region where the southern part of the wave train propagates is simply quiet-Sun region. Therefore, the different magnetic distribution in different regions can account for  the different propagation speeds and morphologies of the different parts of the wave train.

By using the wavelet software, the periodicity of the accompanying pulsation flare is analyzed with the flare lightcurves from radio to hard X-ray waveband. The result indicates that the periods in the pulsation flare ranges from 9 to 1078 seconds, and it is found that all the periods of the QFP wave can be found in the period spectrum of the accompanying pulsation flare. This is in agreement with the previous result that QFP waves and the associated flares share common periods and are possibly caused by a common physical mechanism \citep{liu11,liu12,shen12a,shen13b}. Due to the tight relationship between the QFP wave and the accompanying pulsation flare, the excitation mechanism of the QFP wave should be the same or similar to pulsation flares. Therefore, we can draw on the experience of the excitation mechanism of pulsation flares that have been studied for many years. \cite{nakariakov05} summarized several possible mechanisms for exciting the periodicity of  pulsation flares, including 1) geometrical resonances, 2) dispersive evolution of initially broadband signals, 3) nonlinear processes in magnetic reconnection, and 4) the leakage of oscillation modes from other layers of the solar atmosphere. For the present event, since we do not find evidence supporting the other three mechanism, the periodic nonlinear processes in magnetic reconnection that produces the flare might be important for exciting the QFP wave and the associated pulsation flare. Previous studies have revealed that there are several nonlinear processes in magnetic reconnection that can cause quasi-periodic pulsation in flares. For example, the periodic interaction of the outward moving plasmoids in the current sheet with the ambient magnetic fields \citep[e.g.,][]{kliem00,ni12,yang15,takasao16}, the presence of shear flows in the current layer \citep{ofman06}, and the mechanism of oscillatory magnetic reconnection \citep[e.g.,][]{mclaughlin12a,mclaughlin12b,thurgood17}. In the present event, the periodic radio bursts during the rising phase of the flare could be regarded as the evidence of the periodic releasing of magnetic energy that are possibly caused by the periodic processes in the magnetic reconnection. For other possible exciting mechanisms for QFP waves, \cite{shen12a} found some evidence supporting the scenario that the QFP waves are possibly excited by the leakage of photospheric oscillation to the corona. So far, there is no observational QFP wave events support the possible mechanisms of geometrical resonances and dispersive evolution of initially broadband signals. Therefore, more observational studies based on high resolution data are desirable in the future to understand fully the excitation mechanism and evolution property of QFP waves.

\acknowledgments We thank the observations provided by the {\em SDO}, {\em RHESSI}, and Nobeyama Radioheliograph, and the anonymous referee for his/her valuable suggestions and comments that largely improved the quality of the present paper. This work is supported by the Natural Science Foundation of China (11403097,11633008,11773068), the Yunnan Science Foundation (2015FB191,2017FB006), and the Youth Innovation Promotion Association (2014047) of Chinese Academy of Sciences. The wavelet software was provided by C. Torrence and G. Compo, and is available at URL: \url{http://atoc.colorado.edu/research/wavelets/}.

\begin{sidewaystable}
\caption{Periods of the pulsation flare detected from different observations}
\centering
\begin{tabular*}{24cm}{lm{0.75cm}m{0.75cm}m{0.75cm}m{0.75cm}m{0.75cm}m{0.75cm}m{1.15cm}m{1.15cm}m{1.15cm}m{1.15cm}m{1.15cm}m{1.15cm}m{1.15cm}m{1.15cm}m{1.15cm}}
\hline
\hline
{\bf Wavelength} & \multicolumn{2}{l}{\bf dt\tablenotemark{a} (s)}  &\multicolumn{13}{c}{\bf Period of the flare (s)}\\
\hline
AIA 171 \AA\ & 12 & ... & ... & 25$\pm$3 & ... & ... & ... & ...  & 146$\pm$13 & 188$\pm$16 & 312$\pm$34 & ... & ... &765$\pm$126 & ...\\
AIA 193 \AA\ & 12 &... & ... & 25$\pm$3 & ... & ... & ... & ...  & ... & 164$\pm$35 & 308$\pm$25 & 398$\pm$31 & 522$\pm$43 &796$\pm$150 & ...\\
AIA 211 \AA\ & 12 &... & ... &... & ... & ... & ... & ...  & ... & 180$\pm$35 & ... & 395$\pm$42 & 526$\pm$45 &... & 900$\pm$230\\
AIA 304 \AA\ & 12 &... & ... & ... & ... & ... & ... & 100$\pm$9  & ... & 175$\pm$38 & 263$\pm$20 & 390$\pm$40 & ... &783$\pm$76 & 1049$\pm$130\\
AIA 335 \AA\ & 12 &... & ... & ... & ... & ... & 82$\pm$8 & 120$\pm$20  & ... & 197$\pm$17 & 272$\pm$34 & 391$\pm$32 & 570$\pm$65 &... & 921$\pm$84\\
AIA 94 \AA\ & 12 &... & ... & ... & ... & ... & 91$\pm$10 & ...  & 138$\pm$20 & ... & 333$\pm$38 & ... & ... &758$\pm$65 & ...\\
AIA 131 \AA\ & 12 &... & ... & ... & ... & ... & ...   & 107$\pm$12 & 143$\pm$25 & 218$\pm$25 & 337$\pm$39 & ... &586$\pm$46 & 770$\pm$81 & 1349$\pm$159\\
AIA 1600 \AA\ & 24 &... & ... & ... & ... & ... & ... & 107$\pm$12  & 142$\pm$19 & ... & 294$\pm$25 & ... & ... &... & ...\\
AIA 1700 \AA\ & 24 &... & ... & ... & ... & ... & 75$\pm$8 & 109$\pm$8  & 142$\pm$13 & ... & 299$\pm$32 & ... & ... &... & ...\\
{\em GOES} 1 -- 8 \AA\ & 60 &... & ... & ... & ... & ... & ... & ...  & ... & ... & 300$\pm$28 & ... & 485$\pm$57 & 746$\pm$66 & 1170$\pm$95\\
Nobeyama 1GHz &0.044 & 8$\pm$1 & 20$\pm$2 & 26$\pm$3 & 39$\pm$5 & ... & ... & ...  & ... & ... & ... & ... & ... &... & ...\\
{\em HESSI} 3 -- 6 keV & 4 & 9$\pm$1 & 22$\pm$2 & ... & 38$\pm$4 & 50$\pm$5 & 88$\pm$16 & ...  & ... & 182$\pm$37 & ... & ... & ... &... & ...\\
{\em HESSI} 6 -- 12 keV & 4 & 9$\pm$1 & 22$\pm$2 & ... & 38$\pm$5 & 48$\pm$5 & ... & 110$\pm$18  & ... & 176$\pm$14 & ... & ... & ... &... & ...\\
{\em HESSI} 12--25 keV & 4 & 9$\pm$1 & 22$\pm$2 & ... & 38$\pm$3 & 50$\pm$5 & ... & 111$\pm$18  & ... & 181$\pm$31 & ... & ... & ... &... & ...\\
{\em HESSI} 25--50 keV & 4 & 9$\pm$1 & 22$\pm$2 & ... & 38$\pm$5 & 50$\pm$3 & 78$\pm$10 & 116$\pm$7  & 140$\pm$13 & 189$\pm$15 & 280$\pm$46 & ... & ... &... & ...\\
\hline
 {\bf Average Period}  &  ...  &  9$\pm$1  &  22$\pm$2  &  25$\pm$3  &  38$\pm$4  &  50$\pm$5  &  83$\pm$10  &  110$\pm$13   &  142$\pm$17  &  185$\pm$26  &  300$\pm$32  &  394$\pm$36  &  537$\pm$51  &  770$\pm$94  &  1078$\pm$140 \\
\hline
%\hline
%\multicolumn{2}{l}{\bf Distance to the Flare}  &\multicolumn{13}{c}{\bf Periods of the wave (s)}\\
%\multicolumn{2}{c}{200 Mm (Cut 1)}  &... & ... & 25$\pm$2 & 35$\pm$3 & ... & 75$\pm$10 & 120$\pm$16  &... & 178$\pm$20 & 270$\pm$36 & ... & 554$\pm$105 &... &...\\
%\multicolumn{2}{c}{300 Mm (Cut 2)} & ... & ... &25$\pm$3 &... & 46$\pm$5 & 75$\pm$7 & 106$\pm$12  &... & 160$\pm$18 & 257$\pm$19 & 315$\pm$28 & 445$\pm$59 &... &...\\
%\hline
\end{tabular*}
\tablenotetext{a}{The temporal resolution of the data.}
\tablecomments{Each row lists the detected periods from a certain wavelength data, and the average values for each column are listed in the bottom row. The units of dt and the period are both second.}
\label{tbl2}
\end{sidewaystable}

\end{document}